\documentclass[12pt,preprint2]{aastex}

\def\gs{\mathrel{\raise0.35ex\hbox{$\scriptstyle >$}\kern-0.6em
\lower0.40ex\hbox{{$\scriptstyle \sim$}}}}
\def\ls{\mathrel{\raise0.01935ex\hbox{$\scriptstyle <$}\kern-0.6em
\lower0.40ex\hbox{{$\scriptstyle \sim$}}}}

\newcommand{\um}{\,$\mu$m}

\usepackage{lscape}

\shorttitle{Mid-IR SED simulations}
\shortauthors{Sajina et al.}

\begin{document}

\title{Simulating the {\sl Spitzer} mid-IR color-color diagrams}

\author{Anna Sajina\altaffilmark{1}}
\affil{Department of Physics \& Astronomy, University of British Columbia, \\ Vancouver, BC V6T 1Z1, Canada}
\altaffiltext{1}{Visiting graduate student at the {\sl Spitzer} Science Center for the period during which most of this work was done; sajina@astro.ubc.ca }

\author{Mark Lacy}
\affil{{\sl Spitzer} Science Center, California Institute of Technology, Pasadena, CA 91125, USA}

\author{Douglas Scott}
\affil{Department of Physics \& Astronomy, University of British Columbia, \\ Vancouver, BC V6T 1Z1 Canada}

\begin{abstract}
We use a simple parameterization of the mid-IR spectra of a wide range of galaxy types in order to predict their distribution in the Infrared Array Camera (IRAC) 3.6\um, 4.5\um, 5.8\um, and 8.0\um\ and Multiband Photometer for {\sl Spitzer} (MIPS) 24\um\ color-color diagrams. We distinguish three basic spectral types by the energetically dominant component in the 3--12\um\ regime: stellar-dominated; PAH-dominated; and continuum-dominated. We have used a Markov chain Monte Carlo (MCMC) approach to arrive at a more systematic and robust representation of the mid-IR spectra of galaxies than more traditional approaches. We find that IRAC color-color plots are well suited to distinguishing the above spectral types, while the addition of 24\um\ data allows us to suggest practical 3-color cuts which preferentially select higher redshift sources of specific type. We compare our simulations with the color-color plot obtained by the {\sl Spitzer} First Look Survey (FLS) and find reasonable agreement. Lastly, we discuss other applications as well as future directions for this work.            
\end{abstract}


\section{Introduction}
The 3--12\um\ mid-IR spectral band lies at the interface between the regimes where one observes stellar light directly and where one sees that light reprocessed by dust and gas. The main contributors to galactic emission at mid-IR wavelengths are light from old stellar populations \cite{bos03}, Polycyclic Aromatic Hydrocarbons (PAH) emission features (Puget et al.~1985, Hudgins \& Allamandola~2004), and power law continuum emission, which is, in particular, the dominant component in the mid-IR spectra of type~I AGN mid-IR spectra \cite{clavel00}. This last constribution is likely due to stochastically-heated very small grains \cite{desert}. These sources of emission can be extincted by dust, and further attenuated by discrete absorption features, in particular those from water ice (at $\sim$\,6\um), and Si (at $\sim$\,10\um). Other possible contributors include: Asymptotic Giant Branch (AGB) stars' dust shells; a near sublimation temperature ($\sim$\,1000\,K) blackbody component \cite{lu03}; molecular cooling lines; and nebular lines. Given this complex spectrum, how much {\sl Spitzer} broadband photometry alone can tell us about the nature of the sources, especially if the redshift is also unknown? And how can one interpret the IRAC \cite{faz04} color-color diagram (specifically $S_{5.8}/S_{3.6}$ vs. $S_{8.0}/S_{4.5}$), as well as related diagrams including the MIPS \cite{rieke04} 24\um\ band?

One approach to the above problem would be to use a model with a large number of parameters accounting for all known sources of emission and attenuation; however, this complexity is not justifiable given the small number of available photometric data points. Another approach would be to use a library of fixed spectral templates to choose between, which will allow for the general trends to be studied. However, little information can be extracted from the distribution of observed colors unless a well defined set of templates is used. 

We choose a different approach where we first search for a common empirical description of the $\sim$\,1--12\um\ spectral energy distributions (SEDs) of galaxies, focusing only on what drives the basic shape rather than the detailed structure. In our model, the SED of essentially any galaxy can be representated by a combination of direct stellar emission, PAH emission, and a power law continuum. We fit archival {\sl ISO} spectra of a range of galaxy types with this model using Markov Chain Monte Carlo (MCMC) approach, which provides a handle on the posterior probability distribution of each parameter. In addition, the Markov chains resulting from fitting the spectra of a wide range of galaxy types represent a distribution of potential spectra consistent with the input ones. We simulate the IRAC and IRAC+MIPS color-color diagrams including redshift effects ($z\sim0-2$) by drawing from the generated probability distributions, while including an additional noise term to compensate for the simplicity of the model. 

Lastly, we compare our predictions with early results from the {\sl Spitzer} Space Telescope, and discuss future directions for this work.

\section{The sample}
In order to study how a wide range of galaxy types map onto {\sl Spitzer} photometry bands, we need to use a local sample for which the mid-IR SEDs are determined with sufficiently high spectral resolution and signal-to-noise ratio (SNR).
For this purpose, we use a sample of 60 local galaxies, including ellipticals, normal spirals, ultraluminous infrared galaxies (ULIGs), starbursts, Seyfert galaxies, and HII galaxies (see Table~\ref{sources}). This sample is in no sense complete or free of selection bias (the most obvious of which is toward more active galaxies), our approach being merely to cover as wide a range of spectra as possible. The limiting factor was the availability of 2--12\um\ spectroscopy, especially for classes of galaxy whose volume density in the local Universe is low. We use spectra taken with the PHT-S spectrophotometer (2.5--4.9\um\ and 5.8--11.6\um, with $\lambda/\Delta\lambda$=100--300) on {\sl ISO}, which has a $24\arcsec\times24\arcsec$ aperture. For comparison, {\sl IRAS} had resolution from about 30\arcsec\ at 12\um\ to about 120\arcsec\  at 100\um. We refer to Lu et al.~(2003) for details of the {\sl ISO} spectra and data reduction. We only include in our sample sources detected in all 4 {\sl IRAS} bands, which naturally restricts us to bright, and/or local galaxies ($\left\langle z\right\rangle$=0.009). The typical SNR is $\sim$200\footnote{As estimated by $\sum(x_i/\sigma_i^2)/\sqrt{\sum(1/\sigma_i^2)}$.}, however, some of our sources have significantly higher SNRs which is likely to bias the fitting procedure and certainly makes it difficult to interpret the best $\chi^2$ achieved. To avoid this bias, we have reweighted the spectra with weights corresponding to a signal-to-noise of 200.

The above spectra were extended into the near-IR (NIR) by extracting the flux from 2MASS images using an aperture equivalent to the PHT-S one in a manner similar to that of Lu et al.~(2003)\footnote{The main difference being that we adopted the 2MASS flux calibration from Cohen et al. (2003) without additional offsets. Where available, the images used were taken from the 2MASS Large Galaxy Atlas (Jarrett et al.~2003), and the rest from the 2MASS Extended Source Catalog. }. 

In 10--20\% of the sample the last $\sim$\,0.4\um\ of each half of the PHT-S spectra (i.e.~4.5--4.9\um\ and 11.2--11.6\um) suffer from what appear to be calibration problems, thus for simplicity we ignore these sections in the subsequent analysis. All spectra have been $k$-corrected to redshift zero. 

\section{Model}
\begin{figure}
\centering
\plotone{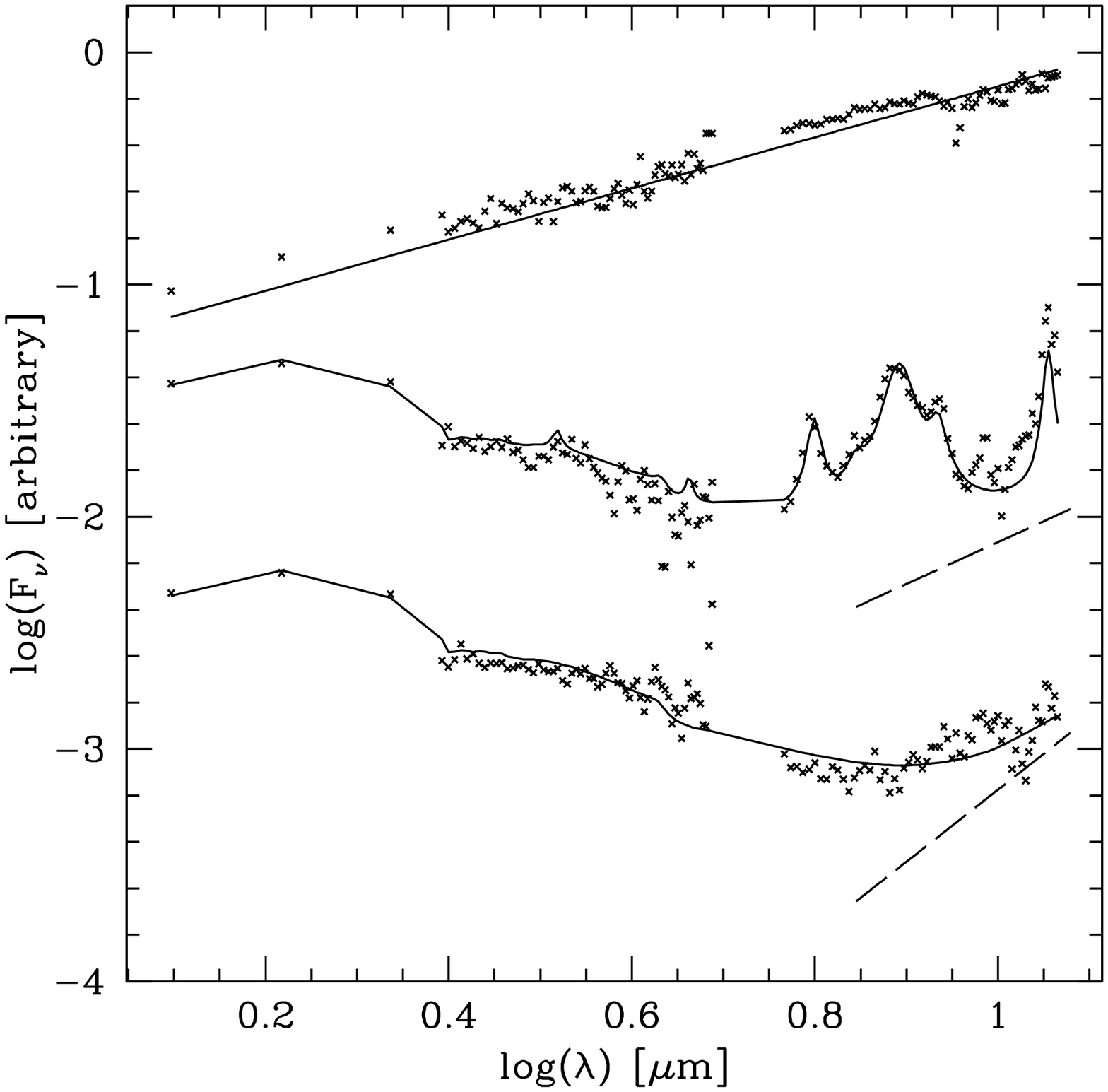}
\caption{\label{types} Here we illustrate the idea behind our model, where the `activity level' of the galaxies increases toward the top. At the bottom is NGC4374 overlaid with the 10\,Gyrs old stellar population template plus a cool continuum ($\alpha\sim3$). In the middle is M51 overlaid with the same as the above only with warmer continuum ($\alpha\sim2$), and our PAH emission template. At the top is IC4329a, which is essentially just a pure power law ($\alpha\sim1$). Note that substructure in the spectra remains; however, these components govern the bulk of the MIR colour variations.}
\end{figure}

As stated in the introduction, the major components governing the shape of the 1--12\um\ SEDs of galaxies are: direct light from old stellar populations; a smooth continuum probably due to very small grains; and PAH emission features. Fig.~\ref{types} illustrates this idea. Our model treats all other possible contributors as noise to the above three. We now discuss each of the three model components in turn. 

{\sl Stars:} We assume a 10\,Gyr old, Salpeter IMF, single stellar population (SSP) which is justifiable since we only consider the SED redward of 1\um\ where it is dominated by K and M giants. At the redshift range we consider, we are likely dominated by evolved systems, and therefore assuming solar metallicity is reasonable. Lower-metallicities would lead to a less pronounced 1.6\um\ bump and 2\um\ break (for a discussion of how these appear to IRAC as a function of redshift see Simpson \& Eisenhardt~1999). We consider the effect of including the dust shell emission of AGB stars in Section\,\ref{agb_shells}. We use the PEGASE2.0 code \cite{fr97} to generate a spectral template with the above specifications. 

{\sl Continuum:} The continuum is given by a flux density $f_{\nu}\propto\nu^{-\alpha}$. Note that, unlike here, a power-law is often used to describe the entire MIR spectrum. It is likely that for the majority of galaxies the origin of this component is the non-thermal emission of very small grains \cite{desert}, although the Wien tail of a $\gs$\,100\,K thermal component is also a possible contributor. Lastly, in some extreme AGN-dominated sources, it can be due to synchrotron emission, although dust tori are a more likely source of MIR emission in typical AGN \cite{pk93}. 

{\sl PAH features:} The PAH bands are modeled as a set of Lorentz profiles\footnote{i.e. $f_{\nu}\propto \{\pi\sigma (1+((\nu-\nu_0)/\sigma)^2)\}^{-1}$} \cite{bbcr98} at 3.3, 6.3, 7.7, 8.6 and 11.3\um. We assume that typically the gap between the 8.6 and 11.3\um\ PAH features accounts fully for the 10\um\ gap, without the need to invoke Si absorption as well \cite{helou}. For the most active galaxies, the usual 7.7\um\ and 8.6\um\  features can be replaced by a broad peak at $\sim$\,8\um. This may be merely continuum bracketed by the water ice and Si absorption features \cite{spoon02}. However, the positions, shapes, and relative strengths of the PAH features can vary in different (particularly extreme) environments and obscuration levels. Thus the PAH-dominated nature of this feature cannot be ruled out \cite{peeters02}. For simplicity, we assume all such features are due to PAH emission, although it is likely that both effects play a role. Detailed modeling of the PAH features is beyond the scope of this work, the template used being merely a fiducial which does not lead to obvious `features' in the residuals of the fits (i.e.~it is some average of the PAH emission features for our sample).

The overall model can be expressed as: $F_{\nu}=a_1\times F_1(\rm{stars})+a_2\times F_2(\rm{PAH})+10^{a_3}\times \nu^{-a_4}$, where the a's designate the free parameters, and the $F$'s are spectral templates. The last two parameters are used to describe the power law component.

\section{Method}
We use a Markov Chain Monte Carlo (MCMC) approach both to find the best-fit set of parameters for our data-set and in order to sample the posterior parameter distributions. We describe only the most relevant details here, but see e.g. Panter et al.~(2003) and Lewis \& Bridle~(2002) for more details of the procedure as well as examples of other applications in astronomy. Starting with uniform priors in our 4 parameters, subsequent trial sets are drawn from a Gaussian around the last accepted set (the proposal distribution, $q$\footnote{Note that the exact shape of the proposal distribution is largely irrelevant, although its `width' in particular affects the efficiency of sampling.}). We follow the Metropolis-Hasting acceptance criterion: $\alpha_{i+1}={\rm min}[u,p_{i+1}q_{i+1}/p_{i}q_{i}]$, where the stochastic element is provided by the random number $u\epsilon[0,1]$.  Since the proposal distribution is symmetric, $q_{i+1}/q_{i}$ is unity. The probability ($p$) that the system finds itself in a given state is given by the Boltzman factor where the `energy' is $\chi^2$ here, and thus $p_{i+1}/p_{i}$ is ${\rm exp}(-\Delta \chi^2/T)$. For a proposal distribution width of 1\% of the total parameter range, $T$\,=\,100 results in $>$\,50\% initial acceptance rate, which constitutes good mixing. We use simulated annealing, i.e. gradually decrease $T$, to allow us to find the high-density regions more efficiently. We run the Monte Carlo for 2000 iterations at each temperature, which meanwhile is decreased as $T_{i+1}$\,=\,0.9\,$T_{i}$\cite{k84} for 20 steps\footnote{These choices are not unique or optimal. We have merely chosen a set of MCMC parameters which work well for this particular case.}. Note that the use of simulated annealing in conjunction with MCMC is non-standard. It has the advantage of vastly speeding up the procedure, and should not severely affect the distribution achieved provided the annealing is slow. Although we have not tested the reliability exhaustively, any deviations will be in the tails of the distribution and therefore irrelevant for the present purposes. There is no universal recipe for deciding when convergence (the `burn-in') is achieved (e.g.~Panter et al.~2003). However, by using a few test cases we find that although it is usually achieved fairly quickly it depends strongly on the initial conditions. We conservatively cut out the first three temperature steps (6000 iterations), which guarantees us convergence in all cases. We then thin the chain by a factor of 30 to decrease the internal correlations which result from every link knowing about its predecessor \cite{lb02}. The result of the MCMC analysis is a chain of parameters (for each of our 60 galaxy SEDs) which properly samples the distribution of best-fit model parameters.   
\section{Results}
\subsection{Individual fits \label{fits}}
\begin{figure}
\centering
\plotone{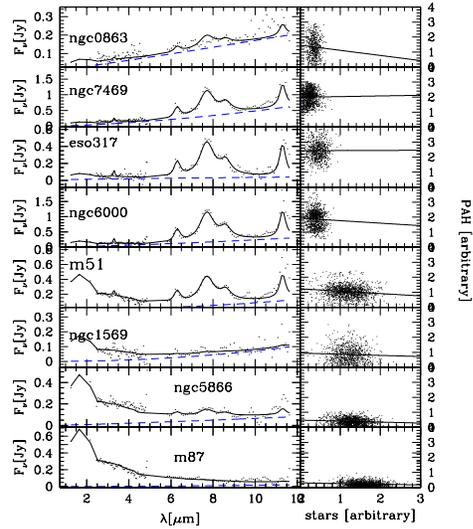}
\caption{\label{mosaic} A sub-sample of our galaxies used to illustrate the continuum of PAH/stars relative strength. The solid line shows the best-fit model, while the dashed line shows the power-law component separately. The right-hand column shows the accepted trials for the stars ($x$-axis) and PAH ($y$-axis) amplitude parameters (each is divided by $\Sigma \nu F_{\nu}$ for ease of comparison) for each galaxy. The scale is the same for all. }
\end{figure}

Fig.~\ref{mosaic} shows the spectra and fits for a sub-sample of our galaxies along with the cross-correlation between the stellar and PAH amplitude parameters ($a_1$ and $a_2$).  Clearly the model provides a reasonable fit to the general shape of the SEDs, although, as expected, residual spectral substructure remains, largely due to variations in the PAH features. This is also seen in the mean of the reduced $\chi^2$s obtained which is $\sim$\,4. 

\begin{figure}
\plotone{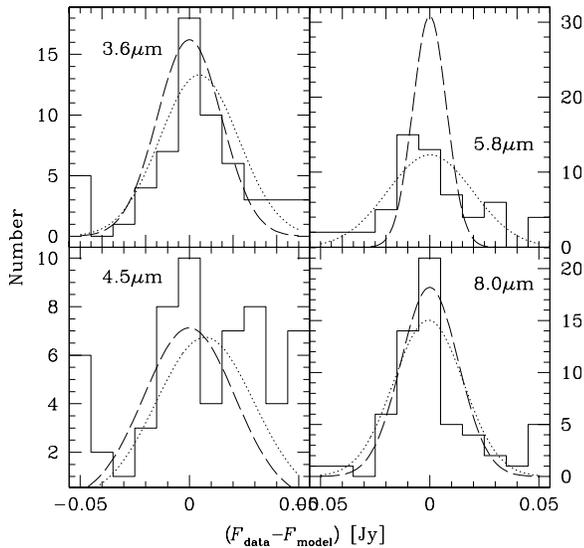}
\caption{\label{erflux} The histograms of the IRAC fluxes obtained from the residuals. Here the dashed line indicates the scatter expected from the uncertainties in the data, while the dotted line is the Gaussian corresponding to the mean and $\sigma$ of the histograms. The positive excess noticeable in the 4.5\um\ plot may be due to the contribution of various strong lines which emit there (e.g. Br\,$\alpha$, CO), although some residual poor calibration can also be the cause, as suggested by the presence of a negative excess as well. The quality of the data does not allow us to distinguish these. The excess scatter in the four bands is: $\sigma_{3.6}$\,=\,10\,mJy, $\sigma_{4.5}$\,=\,8\,mJy, $\sigma_{5.8}$\,=\,18\,mJy, and $\sigma_{8.0}$\,=\,9\,mJy. }
\end{figure}

For our purposes, we need to estimate how much noise to include in the {\sl Spitzer} broadband fluxes in order to account for spectral substructure not included in the model without including the uncertainties associated with the PHT-S spectra themselves. In Fig.~\ref{erflux}, we show the histograms of the difference in IRAC fluxes obtained by convolving the residuals with the IRAC filter profiles as $\Delta F=\int{\Delta S_\nu T_\nu d\nu}/\int{T_\nu d\nu}$, where $T_\nu$ is the transmission curve for the given filter, and $\Delta S_\nu$ is the residual spectrum. These distributions peak near zero showing that our model accounts fairly well for the relevant spectral gradients. We estimate the scatter expected from the uncertainties in the data by taking their average under the filter profiles (analogously to the flux estimates) and dividing by $\sqrt{N}$ (i.e.~the error-in-mean). We compare this scatter with the one obtained and find that the excess is accounted for by $\sigma_{3.6}$\,=\,10\,mJy, $\sigma_{4.5}$\,=\,8\,mJy, $\sigma_{5.8}$\,=\,18\,mJy, and $\sigma_{8.0}$\,=\,9\,mJy. The higher value for $\sigma_{5.8}$ is due to the lower average uncertainty in the data, which is probably related to this filter being at the edge of one of the PHT-S spectral channels. The observed scatter is similar to (rather than twice) that of the other three bands.  For this reason, as well as because using the value obtained results in an unreasonable stretch in the relevant colors, we use $\sigma_{5.8}$\,=\,10\,mJy in the color-color plot simulations (Section~\ref{iracsim}), resulting in uniform scatter across the colors.   
 
\subsection{Correlations \label{correlations}}
The Markov chains constructed above allow us to study how the parameters relate to each other. This is shown in Fig.~\ref{mosaic} where, as expected, the general trend is for strong stellar light emitters to be relatively weak PAH emitters. 

\begin{figure}
\centering
\plotone{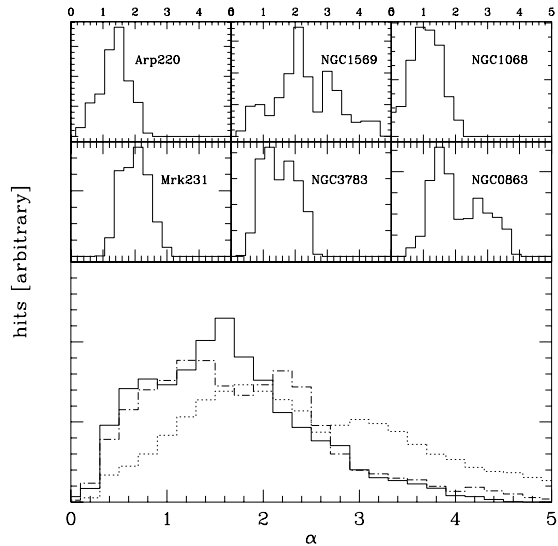}
\caption{\label{cont} The marginalized distributions of the continuum slope $\alpha$. The top six panels show these for a representative sub-sample, while the bottom panel shows the total distribution for the three spectral types, where the solid line is the stellar-dominated distribution, the dotted line is for PAH-dominated, and the dot-dashed line is for continuum-dominated. Note the clear difference in the peaks as well as the typically double-peaked distributions. See Section~\ref{correlations} for discussion.  }
\end{figure}

Fig.~\ref{cont} shows the marginalized Markov chain distributions of the continuum slope, $\alpha$ for a representative sub-sample of six galaxies. In addition, the total distributions of the PAH-dominated, stellar-dominated, and continuum-dominated sources are shown, where the highest peak is, respectively, $\alpha\sim1.8$, $\alpha\sim1.6$, and $\alpha\sim1.2$ for the three types. The last is roughly consistent with the $\alpha=0.84\pm0.24$ obtained by Clavel et al.~(2000) for Seyfert 1 galaxies. The continuum component suffers from some degeneracy with the PAH emission strength, making it difficult to assess what fraction of the width of these distributions is real.

\begin{figure}
\centering
\plotone{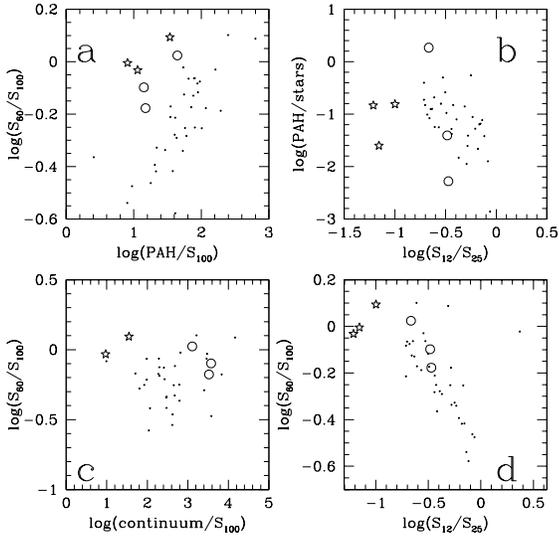}
\caption{\label{corrs} Here PAH/continuum/stars mean the best-fit components integrated over the 3--12\um\ range. Only $z>0.004$ galaxies are used, to minimize aperture bias. Stars mark three SF-dominated active galaxies, while circles show three AGN-dominated active galaxies. See Section~\ref{correlations} for details. }
\end{figure}

For the best-fit models for each of our sample galaxies, we look at how this parameterization relates to the {\sl IRAS} colors. Fig.~\ref{corrs}a shows that the strength of the PAH emission is positively correlated with the $S_{60}/S_{100}$ color, as expected since the later is a measure of the interstellar radiation field (ISRF) which excites the PAH molecules. The exceptions here are the star symbols, which are star-formation-dominated active galaxies (Arp220, NGC4418, IC860), and the circles, which are continuum-dominated AGN (NGC4507, NGC3783, Mrk231). This distinction is seen in Fig.~\ref{corrs}c where the circles are clearly strong continuum emitters (although not the only ones) while the stars are weaker than average continuum emitters (NGC4418 has no discernible continuum). In Fig.~\ref{corrs}d we show the well known {\sl IRAS} color-color plot, and notice that as expected the `stars' occupy the area typically explained by very active galaxies whose PAH carriers are being destroyed by the extreme ISRF. On the other hand, the AGN sources obey the general trend, showing that these colors are insufficient in themselves to distinguish AGN-dominated galaxies, as the lack of PAH emission is compensated by warmer dust resulting in similar $S_{12}/S_{25}$ colors (IRAC colors are far more efficient here, as discussed in Section~\ref{spitzer}). 

Excepting the six galaxies discussed above, the ratio of the PAH to stellar emission appears anti-correlated with the {\sl IRAS} $S_{12}/S_{25}$ color (Fig.~\ref{corrs}c). This cannot be explained through PAH emission, since it affects the 12\um\ filter (which includes the 11.3\um\ and 12.6\um\ features), and thus, if this was the primary driver, the correlation would be positive. It rather suggests that warm ($\sim$\,100\,K) dust which primarily affects the 25\um\ channel, goes along with an increase in the PAH-to-stellar ratio. This makes sense, since both PAH emission and warm dust are expected to trace relatively recent star formation (see e.g.~F\"orster Schreiber et al.~2004) due to their preferentially being excited by energetic UV/optical photons, while the stellar component here is composed of older, redder stars.

\section{IRAC color-color simulation \label{iracsim}}
Since we have 60 independent spectra,  then the overall probability of the data given the model is $P({\bf \theta}|{\bf D},M)=\Pi_{i=1}^{60}P({\bf \theta}|{\bf D}_i,M)$, and we simply link the individual chains obtained above to form a master chain. Our simulation uniformly draws 10,000 times from this chain. It is thus subject to the same selection effects as our original sample. Each set of parameters thus obtained defines an SED through the four-parameter model, at a resolution of $\lambda/\delta\lambda\sim 200$. To obtain the IRAC flux densities from these, we convolve the above with the IRAC filter profiles. To these flux densities we add Gaussian noise, of width equal to the rms for the residuals found in section\,\ref{fits}, in order to account for the missing substructure. Note that this scatter should be regarded as an upper limit as the Markov chain already accounts for some of it and therefore some fraction of the noise is added twice. The resulting IRAC color-color diagram is shown in Fig.~\ref{irac}. Here we distinguish stellar-dominated ({\it red}), PAH-dominated ({\it green}), and continuum-dominated ({\it blue}) sources.

The most prominent feature of this color-color plot is the PAH-to-stars sequence. Since, as discussed previously, this ratio is an indicator of the star-formation activity of a galaxy, this is likely to prove a valuable diagnostic in the study of local sources such as resolved galaxies in the nearby Universe or even galactic star-forming regions. Its exact shape is sensitive to the PAH feature profile and thus can itself be used as a diagnostic for distinguishing for example whether this sequence is significantly different in low metallicity environments than high metallicity ones. As the photometric uncertainty of {\sl Spitzer} observations is typically far smaller than the scatter considered here, such studies are feasible. 

The second aspect is that continuum-dominated sources (AGN) form another sequence with redder $S_{5.8}/S_{3.6}$ colors, for a corresponding  $S_{8.0}/S_{4.5}$  color, than the above. 

Combinations of our three components occupy the region in between these two sequences. Samples with different selection criteria than ours will form the same general shape here, but the density of points across it will differ. 

The use of  mid-IR color-color plots to distinguish star-formation-powered from AGN-powered sources was already known for the {\sl ISO} colors \cite{lau00}. The main difference here is the greater sensitivity to direct stellar light in the IRAC color-color plot.

\subsection{\label{agb_shells} Effect of AGB dust shells}
Assessing the level of contribution of AGB dust shell emission to galaxy spectra is difficult because the effect has the same sense (in terms of the IRAC colors) as interstellar dust. To show this we overplot solar metallicity\footnote{Metallicity has a much smaller effect than age here, since we are mostly sensitive to the relative fraction of the SSP of stars in their AGB phase.} 100\,Myr--10\,Gyr SSP spectra of Piovan et al.~(2003) onto our $z$\,=\,0 color-color plot (Fig.~\ref{irac}). AGBs have a significant effect on the spectra only for relatively young stellar populations. Such populations presumably also emit strongly in their ISM dust (as assumed in our model) which starts to dominate the spectra at $\lambda$\,$>$\,4\,\um. Moreover, in spatially integrated galaxy spectra, the contribution of the young (e.g.~starburst) population can be overpowered in the observed integrated spectra by an older surrounding population. In Fig.~\ref{piovan} we show the actual stellar population SEDs, including AGB dust shell emission and some of the available {\sl ISO} spectra for comparison. We notice that, as expected, the difference is most easily assessed in the roughly 3--5\um\ regime. AGB-dominated SEDs overpredict the power emitted in those wavelengths compared with the near-IR emission. This suggests that the mean age of the stars in the sampled aperture is greater than $\sim$\,0.5\,Gyr. Some of our power-law component may be contributed to by AGB emission in the $>$\,5\um\ spectra (see for example 300\,Myr-old SSP spectra in Fig.~\ref{piovan}). 

For the study of integrated galaxy spectra this effect is not significant. However, it should be kept in mind when looking at finer physical scales inside galaxies, where younger average ages may be observed. In addition, possible foreground from Galactic AGB stars may be an issue specifically near the age track in Fig.~\ref{irac}. 

\subsection{\label{effz} The effect of redshift}
\begin{figure}
\centering
\plotone{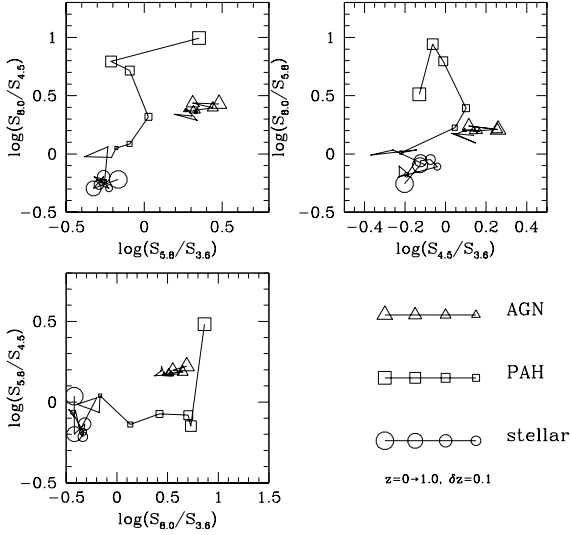}
\caption{\label{zevol1} The $z$\,=\,0\,--\,1 redshift evolution of three model spectra representative of the three major types of mid-IR spectra: continuum-dominated, PAH-dominated; and stellar-dominated. We focus on the ${\rm log}(S_{8.0}/S_{4.5})$ vs.~${\rm log}(S_{5.8}/S_{3.6})$ projection in this work, since it apparently provides the best separation between the PAH-dominated and continuum-dominated sources in particular.}
\end{figure}

We show in Fig.~\ref{zevol1} the $z$\,=\,0\,--\,2 redshift evolution in the IRAC color-color plot for models representative of the three basic types discussed in this work. We can understand the general trends of movement in the color-color diagrams through considering the main PAH features moving through the IRAC bands. 

 For PAH-dominated sources, the $S_{5.8}/S_{3.6}$ color shifts blueward between $z$\,$\sim$\,0\,--\,0.1, due to the 6.2\um\ feature leaving the 5.8\um\ filter. By $z\sim0.2$ the 3.3\um\ feature leaves the 3.6\um\ filter and thus the color shifts slightly redward again, until the stellar 1.6\um\ bump begins to enter the 3.6\um\ filter past $z\sim0.7$. The color shifts redward strongly when the 5.8\um\ filters begins to sample the stellar peak as well, which makes this color a sensitive redshift estimator past $z$\,$\sim$\,1 \cite{lf04}. On the other hand, the $S_{8.0}/S_{4.5}$ color starts off red, due to the powerful 7.7\um\ PAH feature complex, and gradually turns bluer, due to the influence of the 3.3\um\ feature on the 4.5\um\ filter ($z\sim0.2-0.5$) coupled with the 6.2\um\ feature replacing the 7.7\um\ one. Past $z\sim0.5$ the strongest PAH features leave the 8\um\ filter altogether, and at this point star-forming galaxies enter the general vicinity of $z\sim0$ stellar-dominated sources, which is also where contamination from foreground stars (especially dust enshrouded ones) is most severe. Thus, these two colors alone are insufficient to distinguish the high-$z$ starforming galaxies.  

Stellar-dominated sources occupy the same general vicinity until $z\sim0.6$ when the 1.6\um\ bump enters the 3.6\um\ filter, while further redshifting (especially past $z\sim2$) causes a red $S_{5.8}/S_{3.6}$ color due to their red optical continua probably leading to a sequence similar to the `AGN-sequence', however offset in its $S_{8.0}/S_{4.5}$ color. 

Continuum-dominated sources, being relatively featureless, also do not leave the general vicinity they occupy locally, instead they just shift along the `continuum-sequence' due to slight changes in slope. Only largely unobscured AGN will ever display `blue' IRAC colors when the restframe optical wavelengths begin to be sampled at $z>2$.

 \subsection{The use of 24\um\ data at $z$\,$>$\,1}
 Due to the lack of available data, we do not formally extend our model past 12\um. However, as seen in the previous section, the diagnostic power of the 8\um\ PAH feature complex becomes inaccessible to the IRAC colors alone past $z$\,$\sim$0.7. We overcome this by the use of the MIPS 24\um\ filter once our 1--12\um\ spectra are redshifted into it at $z$\,$\gs$\,1. As discussed previously, the narrow PAH-to-stars sequence is likely to be an important diagnostic arising from the IRAC color-color plot. Here we consider the lg($S_{24}/S_{5.8}$) color which at $z$\,$>$\,1 mimicks the lg($S_{8}/S_{4.5}$) local PAH-to-stars sequence. In Fig.~\ref{color_cut}, we show the redshift dependence of the {\it three} colors for the different spectral types. Obviously, the use of such an implicit `24\um-detection' criterion is also useful in removing foreground contamination, which is an issue in particular as the IRAC colors of high-$z$ galaxies can be very `star'-like (see Section~\ref{effz}). 

\subsection{\label{colsel} Color selection}  
In Fig.~\ref{zevol}, we present specific color-cuts in the three colors discussed ($S_{5.8}/S_{3.6}$, $S_{8.0}/S_{4.5}$, and $S_{24.0}/S_{5.8}$), which preferentially select sources with specific types of mid-IR spectra at various redshift ranges. As in the local IRAC color-color plot, the most prominent feature here is the PAH-to-stars sequence, which shifts left with redshift, allowing both for a handle on the underlying $N(z)$ and some sense of the physical characteristics of the sample (since the above sequence is most likely a tracer of the SF activity of a galaxy). Since the various regimes discussed can overlap significantly, this is likely to be most useful for large samples where `mis-hits' are washed out, while for small samples (order of 10) such color-cuts are likely to be hard to interpret unless additional information is available.

\section{Discussion}
\subsection{\label{spitzer} Comparison with {\sl Spitzer} results}
The First Look Survey (FLS), conducted with the IRAC and MIPS instruments on the {\sl Spitzer} Space Telescope, is a publicly available, shallow, 4\,deg$^2$ survey\footnote{http://ssc.spitzer.caltech.edu/fls/}. The IRAC 4-band catalog \cite{mark04} has flux density limits (5\,$\sigma$ in a 5\,$^{\prime\prime}$ aperture) of $\sigma_{3.6}$\,$\approx$\,7\,$\mu$Jy, $\sigma_{4.5}$\,$\approx$\,8\,$\mu$Jy, $\sigma_{5.8}$\,$\approx$\,60\,$\mu$Jy, and $\sigma_{8.0}$\,$\approx$\,50\,$\mu$Jy. We use this catalog to explore qualitatively how our model compares with real data.  Fig.~\ref{irac_z0.2} shows the FLS data on the top, with a simulation of it from our model shown below. To create the latter we use the master chain discussed previously. The $N(z)$ used covers $z=0.2-2.0$ and is a combination of two Gaussians centered at $z$\,=\,0.2 and $z$\,=\,0.6 of widths 0.08 and 0.3 respectively. Note that this is merely a curve giving a reasonable approximation to the observed FLS IRAC color-color plot. Work is in progress to more robustly derive the redshift distributions per luminosity per spectral type by comparing our simulations with observed color-color distributions (see following section). In addition, since stars have not been removed from the FLS catalog, we add a Gaussian distribution centered at (--0.4, --0.6) with $\sigma$=0.15 and the same centered at (--0.3, --0.5) to cover the gap from the above to the old stellar population colors used in our model, and thus emulate the expected continuum of colors from different types of stars (no attempt has been made to optimize this).  Although not perfect in detail, this nonetheless shows that the essential elements of the observed FLS color-color plot are well reproduced by our model. 

Our model suggests, that the two `bunny ears' in the FLS color-color plot separate the $z\sim0-1$ continuum-dominated sources (right ear) from the $z\sim0.1-0.3$ PAH-dominated sources (left ear). This ability to separate out the continuum-sources (i.e. AGN) based on {\sl Spitzer} photometry, which is far less affected by extinction than optical observations, was explored by Lacy et al. (2004a) in order to assess the fraction of AGN-dominated sources missed by optical surveys due to dust obscuration. 

We also want to compare our simulations with the properties of known $z>1$ sources observed with IRAC and MIPS. Ivison et al.~(2004) present the {\sl Spitzer} colors of nine MAMBO sources. The 1200\um\ MAMBO filter, by virtue of its negative $k$-correction (climbing the Rayleigh-Jeans tail of the thermal dust emission peak), selects higher redshift sources. Ivison et al.~(2004) indicate how the {\sl Spitzer} colors can be used to separate out AGN from star-formation(PAH)-dominated sources. We found that, although our sample on average did not require extinction applied to it to achieve acceptable fits\footnote{As we are only concerned with $\lambda>1$\um, modest extinction cannot be excluded.}, this was essential to explain the above observations. In Fig.~\ref{mambo_sources} we show the Ivison et al.~(2004) color plot with extinction gradually applied (using the Draine et al.~2003 Milky Way $R_V$\,=\,3.1 extinction curve) to our simulations for the case of $z=1$ and $z=2$. We find that although AGN sources preferentially lie on the AGN-sequence defined by Ivison et al. based on Mrk231, it is more acurately described as an extinction sequence at high ($z$\,$\gs$\,2) redshifts. PAH-dominated and continuum-dominated (i.e.~AGN) sources are found on this sequence with roughly equal probability, provided relatively high extinction levels (e.g.~$\tau_V\sim10$) are applied to both. We also note that the use of this plot for even rough redshift estimates is complicated by the large variations arising as the PAH 7.7\um\ feature complex enters and leaves the 24\um\ filter, meaning that high values of $S_{24}/S_8$ do not necessarily indicate low redshift. 

\subsection{Future work}
In this work we have made use of the best mid-IR data available to date over a wide range of galaxy types. However, the Infrared Spectrograph (IRS) on {\sl Spitzer} will soon improve on these, both in data quality and sample size. IRS does not observe the $\lambda$\,$<$\,5\um\ range (i.e.~it will miss most of the direct stellar emission), but it does extend to 38\um, which will allow us to include additional components to our model such as warm dust emission. We plan to make use of these future {\sl Spitzer} data in order to repeat this analysis in a complementary manner.   

Further, we plan to extend this approach to modeling galaxy SEDs to longer wavelengths in particular far-IR/sub-mm. The far-IR/sub-mm is, by comparison with the mid-IR, relatively simple and can be modeled empirically with only 1-2 parameters as shown for the {\sl ISO} key project galaxies by Dale et al.~(2001). The observed correlations between our model parameters and the {\sl IRAS} colors suggest that such an extension is feasible\footnote{An additional indication that such an extension arises naturally is the observation that the PAH and sub-mm emission of resolved galaxies correlate fairly strongly \cite{hb02}.}.
      
The full potential of the MCMC approach has not yet been exploited. In particular, work is in progress to assess to what degree we can estimate the $N(z)$, luminosity functions and so on of observed distributions, which is only hinted at in Fig.~\ref{irac_z0.2}. For such a study, a more complete investigation of the effect of simulated annealing on the resulting distribution as well as potential biases and numerical optimizations will need to be addressed. 
 
Lastly, we would like to understand further how this parameterization of galaxies' SEDs relates to physical properties such as star-formation rate, dust properties, and metallicity. We plan to approach this primarily through detailed UV/optical-to-MIR SED modeling of local, resolved galaxies.
                     
\acknowledgements
This work uses {\sl ISO} archive data, specifically galaxy PHT-S spectra reduced and graciously provided to us by N.\,Lu and M.\,Hur of the Infrared Processing and Analysis Center (IPAC). We are also grateful to Lorenzo Piovan for providing us with his SSP spectra, accounting for the contribution of AGB dust shells. 
AS was supported by a Visiting Graduate Student fellowship at the {\sl Spitzer} Science Center for the period during which most of this work was done. She is grateful for conversations with many people there, which helped shape the direction of this project. This work was supported in part by the Natural Science and Engineering Research Council of Canada (NSERC). This research has made use of the NASA/IPAC Extragalactic Database (NED) which is operated by the Jet Propulsion Laboratory, California Institute of Technology, under contract with the National Aeronautics and Space Administration (NASA). This work has made use of data obtained as part of the Two Micron All Sky Survey (2MASS), a joint project between the University of Massachusetts and IPAC, funded by NASA and the National Science Foundation.

\clearpage

\begin{figure}
\caption{\label{irac} Here we show the Monte Carlo generated IRAC color-color diagram, where red indicates starlight-dominated, green is PAH-dominated, and blue designates continuum-dominated. The tracks show the effect of AGB dust shells on the IRAC colors for a range of stellar population ages (given in log[Myr]). }
\end{figure}

\begin{figure}
\plotone{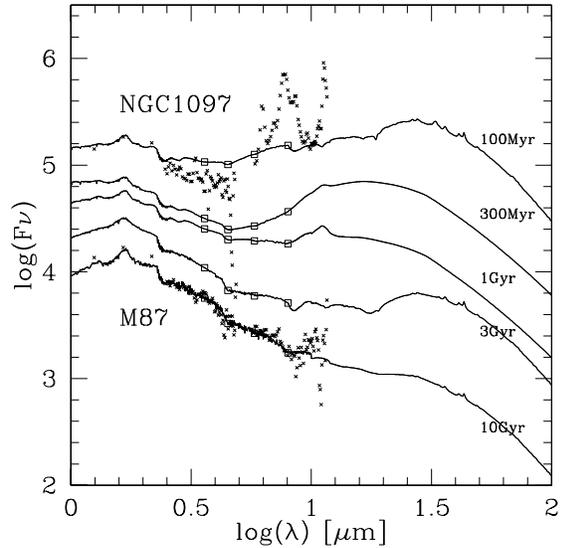}
\caption{\label{piovan} The SSP spectra \cite{piovan03} corresponding to the age track in Fig.~\ref{irac} overlaid with representative galaxy spectra. See Section~\ref{agb_shells} for discussion. }
\end{figure}

\begin{figure}
\caption{\label{color_cut} The color evolution of the stellar-dominated, continuum-dominated, and PAH-dominated sources as a function of redshift. Note in particular the influence of the 7.7\um\ PAH feature at $z\ls0.5$, and the 1.6\um\ peak at $z\gs1.3$. The scatter represents $\pm$\,2\,$\sigma$. See Section~\ref{effz} for further details. }
\end{figure}

\begin{figure}
\centering
\caption{\label{zevol} The $z$\,=\,0\,--\,2 color evolution based on our simulations. The color scheme is as in Fig.~\ref{irac}. The top panels are divided into regions which preferentially select continuum-dominated sources at $z\sim0-2$ (region 1), PAH-dominated sources at $z\sim0.05-0.3$ (region 2), stellar- and PAH-dominated sources at $z$\,$\sim$\,0.3\,--\,1.6 (region 3), and same at $z$\,$\gs$\,1.6 (region 4). 
The bottom panels show simulations in the range $z$\,=\,1\,--\,2 only. The $z$\,=\,0 region marks the spread of colors for our sample, while the  $z$\,$<$\,1 is our best guess based on the assumption that the lg($S_{24}/S_{5.8}$) will not drop below this range before the 24\um\ filter begins to sample the 10\um\ dip at $z$\,$\sim$\,1.5. Stellar-dominated systems here are off-scale (see Fig.~\ref{color_cut}). See Section~\ref{colsel} for a discussion.}
\end{figure}

\begin{figure}
\centering
\vspace*{10cm}
\leavevmode
\caption{\label{irac_z0.2} {\it top:} The FLS IRAC color-color plot. {\it bottom:} A simulation from our model. See Section~\ref{spitzer} for details and discussion. }
\end{figure}

\begin{figure}
\centering
\caption{\label{mambo_sources} The effect of extinction. Here as usual green indicates PAH-dominated sources and blue is continuum-dominated sources. The red filled circles are the MAMBO sources from Ivison et al.~(2004). }
\end{figure}

\begin{deluxetable}{lccccc} 
\tabletypesize{\footnotesize}
\tablecolumns{7} 
\tablewidth{0pc} 
\tablecaption{\label{sources} Basic data for our sample } 
\tablehead{    
\colhead{Name} & \colhead{RA[J2000]}   & \colhead{Dec[J2000]}    & \colhead{Class\tablenotemark{a}} & 
\colhead{$z$}    & \colhead{$\rm{log}$$(S_{60}/S_{100})$}}
\startdata 
Arp220 & 15:34:57.28 & 23:30:11.3 & LINER, HII, Sy2 & 0.01813 & --0.03 \\
CenA & 13:25:27.50 & --43:01:10.9 & S0, Sy2 & 0.00183 & --0.35 \\
ESO317-G23 & 10:24:42.50 & --39:18:21.1 & SBa & 0.00965 & --0.28 \\
IC860 & 13:15:03.53 & 24:37:07.9 & Sa, HII & 0.01116 & --0.00 \\
IC883 & 13:20:35.34 & 34:08:22.0 & Im, HII, LINER & 0.02335 & --0.21 \\
IC1459 & 22:57:10.46 & --36:27:44.5 & E3, LINER & 0.00564 & --0.36  \\
IC4329a & 13:49:19.15 & --30:18:34.24 & SA0+, Sy1.2 & 0.01605 & 0.09  \\
IC4595 & 16:20:44.26 & --70:08:35.2 & Sc & 0.01137 & --0.42 \\
M51 & 13:29:52.66 & 47:11:42.6 & SAbc, HII, Sy2.5 & 0.00154 & --0.63   \\
M81 & 09:55:33.23 & 69:03:55.0 & SAab, LINER, Sy1.8 & 0.000133 & --0.67  \\
M83 & 13:37:00.46 & --29:51:55.7 & SABc, HII, Sbrst & 0.00172 & --0.33  \\
M87 & 12:30:49.51 & 12:23:28.6 & E, NLRG Sy & 0.00436 & --0.02 \\
Mrk231 & 12:56:14:18 & 56:52:25.26 & SAc, Sy1 & 0.04217 & 0.02  \\
Mrk331 & 23:51:26.24 & 20:35:08.2 & S?, HII, Sy2 & 0.01848 & --0.06  \\
Mrk359 & 01:27:32.2 & 19:10:39.5 & SB0a, Sy1.5 & 0.01739 & --0.19  \\
Mrk520 & 22:00:41.43 & 10:33:07.5 & Sb, Sy1.9 & 0.02661 & --0.21  \\
Mrk817 & 14:36:22.08 & 58:47:39.0 & SBc, Sy1.5 & 0.03145 & --0.03  \\
NGC0701 & 01:51:03.71 & --09:42:10.2 & SBc, Sbrst & 0.00610 & --0.34  \\
NGC0863 & 02:14:33.54 & --00:46:00.4 & SAa, Sy1.2 & 0.02638 & --0.48  \\
NGC1022 & 02:38:32.47 & --06:40:39.0 & SBa, HII, Sbrst & 0.00485 & --0.13  \\
NGC1056 & 02:42:48.34 & 28:34:26.8 & Sa, HII, Sy & 0.00515 & --0.25  \\
NGC1068 & 02:42:40.83 & --00:00:47.0 & SAb, Sy1, Sy2 & 0.00379 & --0.10  \\
NGC1097 & 02:46:19.06 & --30:16:28.0 & SBb, Sy1 & 0.00425 & --0.28  \\
NGC1125 & 02:51:39.51 & --16:39:08.4 & SAB, Sy2 & 0.03175 & --0.07  \\
NGC1222 & 03:08:56.81 & --02:57:17.6 & S0, HII & 0.00818 & --0.06  \\
NGC1326 & 03:23:56.40 & --36:27:49.6 & SB0/a, LINER & 0.00454 & --0.25  \\
NGC1377 & 03:36:39.01 & --20:54:06.2 & S0, HII & 0.00598 & 0.10  \\
NGC1385 & 03:37:28.22 & --24:30:04.0 & SBcd & 0.00498 & --0.33  \\
NGC1546 & 04:14:36.60 & --56:03:38.9 & SA0+? & 0.00426 & --0.58  \\
NGC1569 & 04:28:11.14 & 64:50:52.25 & IBm, Sbrst, Sy1 & 0.000244 & --0.02  \\
NGC3227 & 10:23:30.60 & 19:51:55.2 & SAB, Sy1.5 & 0.00386 & --0.35  \\
NGC3705 & 11:30:05.82 & 09:16:36.3 & SABab, LINER, HII & 0.00340 & --0.46  \\
NGC3783 & 11:39:01.78 & --37:44:19.6 & SBa, Sy1 & 0.00973 & --0.18  \\
NGC3949 & 11:53:41.42 & 47:51:31.7 & SAbc, HII & 0.00267 & --0.38  \\
NGC4027 & 11:59:30.59 & --19:15:48.1 & SBdm & 0.005574 & --0.39  \\
NGC4051 & 12:03:09.58 & 44:31:52.9 & SABbc, Sy1.5 & 0.00234 & --0.53  \\
NGC4102 & 12:06:23.37 & 52:42:41.2 & SABb?, HII, LINER & 0.00282 & --0.17  \\
NGC4194 & 12:14:10.10 & 54:31:41.9 & IBm, BCG, HII  & 0.00836 & --0.08  \\
NGC4253 & 12:18:26.43 & 29:48:46.0 & SBa, Sy1.5 & 0.01293 & --0.06  \\
NGC4365 & 12:24:28.87 & 07:19:04.6 & E3 & 0.00415 & --1.17  \\
NGC4374 & 12:25:03.09 & 12:53:10.5 & E1, LERG, LINER & 0.00354 & --0.37  \\
NGC4418 & 12:26:54.65 & --00:52:40.5 & SABa, Sy2 & 0.00727 & 0.09  \\
NGC4490 & 12:30:36.88 & 41:38:23.4 & SBd & 0.00188 & --0.25  \\
NGC4507 & 12:35:36.67 & --39:54:33.16 & SABab, Sy2 & 0.01180 & --0.10  \\
NGC4569 & 12:36:49.78 & 13:09:46.2 & SABab, LINER, Sy & 0.002148 & --0.50  \\
NGC4593 & 12:39:39.35 & --05:20:38.8 & SBb, Sy1 & 0.00900 & --0.29  \\
NGC4691 & 12:48:13.47 & --03:19:59.1 & SB0/a, HII & 0.00137 & --0.15  \\
NGC5253 & 13:39:55.75 & --31:38:30.8 & Im, HII, Sbrst & 0.00135 & --0.011  \\
NGC5433 & 14:02:36.03 & 32:30:37.5 & Sdm & 0.01452 & --0.25  \\
NGC5548 & 14:17:59.48 & 25:08:12.11 & SA0/a, Sy1.5 & 0.01717 & --0.18  \\
NGC5866 & 15:06:29.41 & 55:45:47.2 & S0, HII/LINER & 0.00224 & --0.57  \\
NGC5915 & 15:21:33.16 & --13:05:32.5 & SBab & 0.00758 & --0.17  \\
NGC5962 & 15:36:31.70 & 16:36:31.8 & SAc, HII & 0.00653 & --0.42  \\
NGC6000 & 15:49:49.28 & --29:23:13.1 & SBbc, HII & 0.00732 & --0.17  \\
NGC6753 & 19:11:23.29 & --57:02:55.4 & SAb & 0.01042 & --0.46  \\
NGC7218 & 22:10:11.71 & --16:39:35.7 & SBc & 0.00554 & --0.34  \\
NGC7418 & 22:56:35.90 & --37:01:45.7 & SABcd & 0.00482 & --0.54  \\
NGC7469 & 23:03:15.58 & 08:52:25.9 & SABa, Sy1.2 & 0.01632 & --0.13  \\
NGC7592 & 23:18:22.10 & --04:24:59.8 & LINER, Sy2 & 0.02444 & --0.12  \\
\enddata 
\tablenotetext{a}{The NASA Extragalactic Database (NED) morphological and spectroscopic classes.}
\end{deluxetable}

\end{document}